\documentclass[]{aa}  
\usepackage{graphicx}
\usepackage{txfonts}
\usepackage{subfigure}
\usepackage{natbib}
\usepackage{longtable}
\usepackage{comment}
\usepackage{hyperref}

\usepackage{ulem}

\usepackage{color}
\usepackage{amstext}

\begin{document}


\def\etal{et al.$\;$}
 

\def\kms{km\thinspace s$^{-1}$}
\def\Lsun{L$_\odot$}
\def\Msun{M$_\odot$}
\def\ms{m\thinspace s$^{-1}$}
\def\percc{cm$^{-3}$}
\title{The formation radius of HCN in O-rich asymptotic giant branch stars}

\titlerunning{Formation radius of HCN in O-rich AGB stars}

\subtitle{}

\author{L. Marinho
        \inst{1}
        \and
        J. P. Fonfría
        \inst{1}
        \and
        M. Agúndez
        \inst{1}
        \and
        L. Velilla-Prieto
        \inst{1}
        \and
        G. Quintana-Lacaci
        \inst{1}
        \and
        J. Cernicharo
        \inst{1}
}

\institute{Instituto de Física Fundamental, CSIC, C/ Serrano 123, 28006 Madrid, Spain        \email{louise.marinho@iff.csic.es}
}

\date{Accepted in A\&A 16th June 2026}

\abstract
  {Molecules detected in circumstellar envelopes around asymptotic giant branch (AGB) stars are generally well explained in terms of formation under chemical equilibrium in the stellar atmosphere or photochemistry in the outer expanding layers. However, several molecules are detected in the inner circumstellar regions with abundances orders of magnitude above the predictions of chemical equilibrium. Anomalously abundant molecules comprise H$_2$O, NH$_3$, SiH$_4$, and PH$_3$ in C-rich envelopes and HCN, CS, and NH$_3$ in O-rich sources. These molecules are formed by some yet unknown nonequilibrium processes.}  
   {We aim to shed light on the nonequilibrium process that enhances the abundance of anomalously abundant molecules in the inner regions of circumstellar envelopes. Here, we focus on HCN in O-rich stars to constrain its formation radius.}
   {We observed the O-rich AGB stars R\,Crt and IK\,Tau with ALMA at high angular resolution (HPBW=~0.05") in the $J$\,=\,$4-3$ line of HCN in both the ground and the $\nu_2$\,=\,1 vibrational states. The radial distribution of the emission was modeled using a large velocity gradient radiative transfer code in which we considered an abundance profile with a central hole.}
   {The models with a hole in the abundance distribution of HCN reproduce the specific shape of the radial emission distribution around the two studied O-rich stars. The best models locate the formation radius of HCN at 4-6 $R_\star$ in R\,Crt and 3-5 $R_\star$ in IK\,Tau.}
   {The formation radius of HCN inferred from observations is consistent with models that invoke shocked-induced chemistry but also with models where photochemistry acts as the solely disequilibrium process. High angular resolution observations of more O-rich stars and tailored shocked-induced and/or photochemical models are needed to unambiguously unveil the underlying mechanism behind anomalously abundant molecules in AGB envelopes.}
   
   \maketitle
   \nolinenumbers


\section{Introduction}
\label{sec:int}

Evolved stars provide strong mechanical, chemical, and radiative feedback on their host environment \citep{Habing1996,HofnerOlofsson2018}. During the asymptotic giant branch (AGB) phase, stars lose mass at a high rate, creating extended circumstellar envelopes (CSEs) composed of molecular gas and dust grains. In general terms, molecules in CSEs are either formed in the warm and dense inner regions, where chemical equilibrium prevails \citep{Tsuji1973, Agundez2020}, or in the outer expanding layers due to the action of photochemistry driven by external ultraviolet (UV) photons \citep{Agundez2010B, Glassgold1986, Nejad1987, Millar2000}. 

Chemical equilibrium naturally explains the marked differentiation between O- and C-rich envelopes as a consequence of the C/O ratio at the stellar surface being below or above unity. Thus, around O-rich stars (C/O\,$<$\,1) O-bearing parent molecules, such as H$_2$O, SiO, SO, SO$_2$, PO, and CO$_2$, are usually found \citep{Tsuji1997,Ryde1998, Gonzalez-Delgado2003, Maercker2016, Ziurys2018, Danilovich2020,Fonfria2020, Massalkhi2020, Baudry2023, Wallstrom2024}, while carbon-bearing molecules, such as C$_2$H$_2$, CH$_4$, C$_2$H$_4$, C$_4$H$_2$, C$_6$H$_2$, HCN, CS, SiC$_2$, and HCP are commonly present around carbon-rich stars \citep{Keady1993,Cernicharo2001,Agundez2007,Fonfria2008,Fonfria2017,Fonfria2018,Schoier2013,Massalkhi2018,Massalkhi2019,Menten2018,Unnikrishnan2025}. However, some molecules are observed with abundances orders of magnitude above the predictions from chemical equilibrium \citep{Agundez2020}. The most remarkable cases of anomalously abundant molecules in C-rich envelopes are H$_2$O, NH$_3$, SiH$_4$, and PH$_3$ \citep{Decin2010_h2o,Neufeld2011,Neufeld2011_irc10216,Schmidt2016,Agundez2014,Keady1993}, while HCN, CS, and NH$_3$ stand out in O-rich objects \citep{Agundez2010B, Schoier2013, Danilovich2019, Massalkhi2020, Menten2010, Wong2018}. These molecules are formed in the inner circumstellar layers, and thus their overabundance with respect to chemical equilibrium suggests that there must be important nonequilibrium processes at work in these regions.

Two different types of processes have been proposed to account for the departure from chemical equilibrium in the inner layers of CSEs. On the one hand, shocks resulting from the periodic pulsation of the star \citep{Bowen1988,HofnerOlofsson2018} can induce significant changes in the chemical composition \citep{Willacy1998,Duari1999,Cherchneff2006,Cherchneff2011,Cherchneff2012,Gobrecht2016}. On the other hand, UV radiation penetrating into the inner layers through the clumpy structure of the envelope or arising from binary companions or chromospheric activity can also drive significant departures from chemical equilibrium via photochemistry \citep{Decin2010_h2o,Agundez2010,VandeSande2018,VandeSande2019,VandeSande2022}. Internal X-rays can also drive the composition out of equilibrium, as for HCO$^+$ and HNC, which are enhanced in the inner regions of O-rich AGB envelopes \citep{Alonso2025}. Whether one or a combination of these processes is at the origin of the anomalously abundant molecules observed in the inner regions of CSEs remains unclear. 

Whatever nonequilibrium process is at work, the departure from chemical equilibrium must occur in the very inner layers. Therefore, constraining the spatial distribution of anomalously abundant molecules in these regions, and in particular determining the exact location where they are formed, is of great interest to shed light on the underlying formation mechanism. 

In this study, we present ALMA high angular resolution observations of HCN around two O-rich AGB stars (R\,Crt and IK\,Tau), from which it is possible to constrain the formation region of this anomalously abundant molecule. In Sect.\,\ref{sec:obse} we present the ALMA observations. In Sect.\,\ref{sec:meth} we describe the radiative transfer model. In Sect.\,\ref{sec:result} we report our results, discuss them in Sect.\,\ref{sec:disc}, and present our main conclusions in Sect.\,\ref{sec:Conclusions}.

\section{Observations}
\label{sec:obse}

\begin{figure*}
\centering
\includegraphics[width=0.40\textwidth,clip]{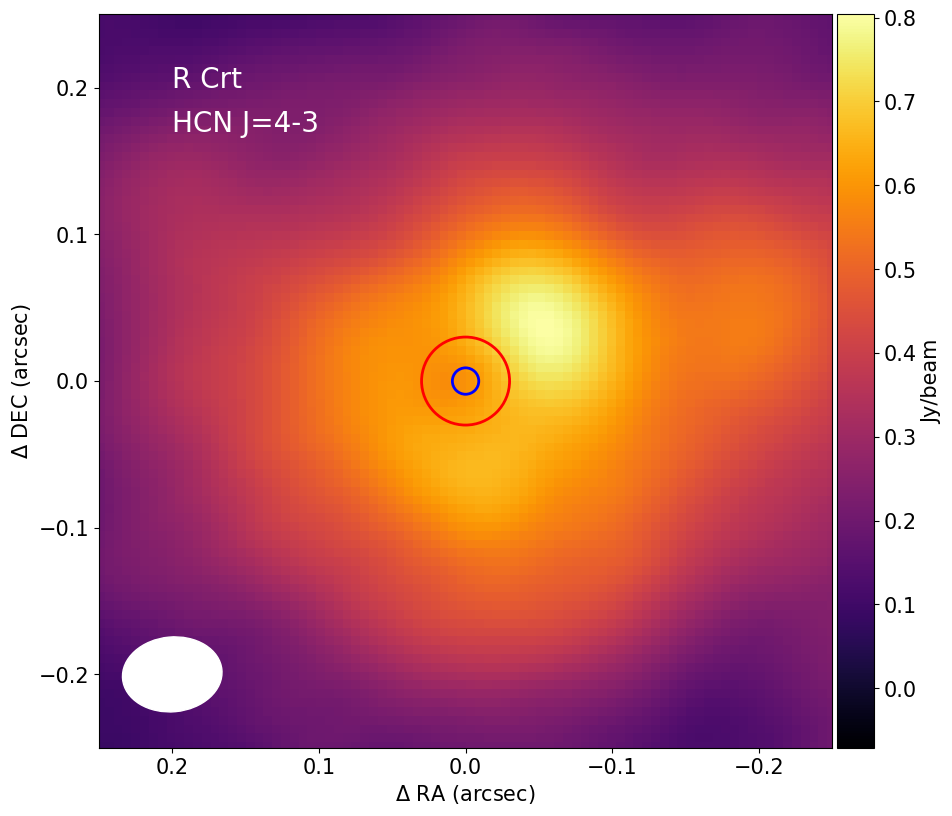}
\includegraphics[width=0.40\textwidth,clip]{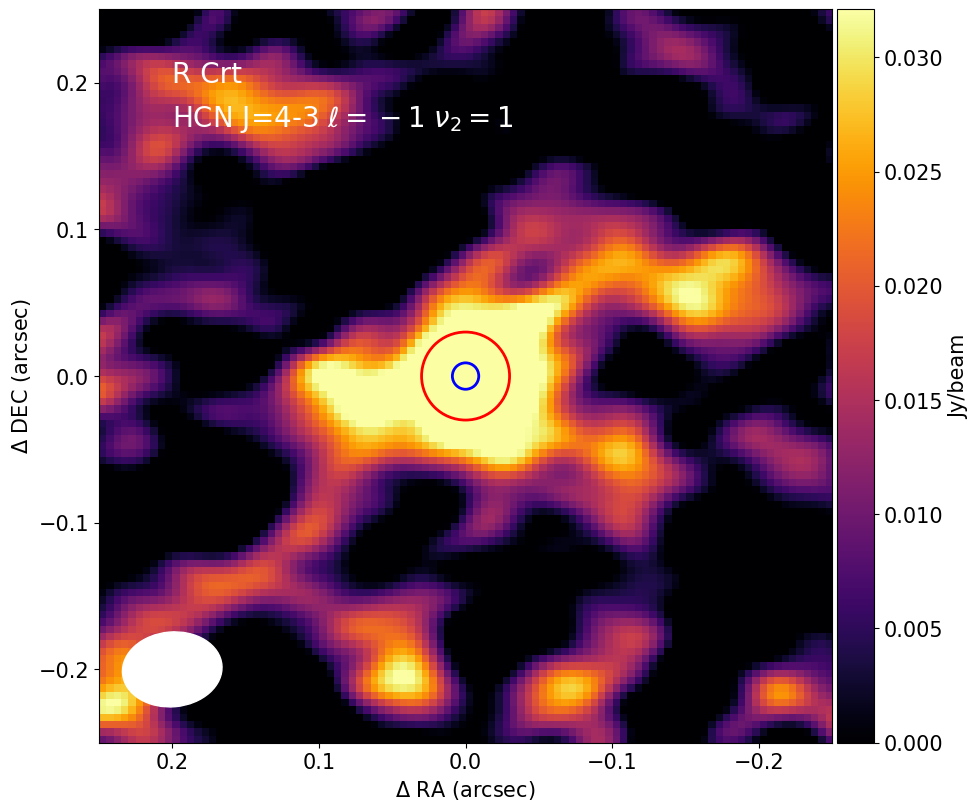}
\includegraphics[width=0.40\textwidth,clip]{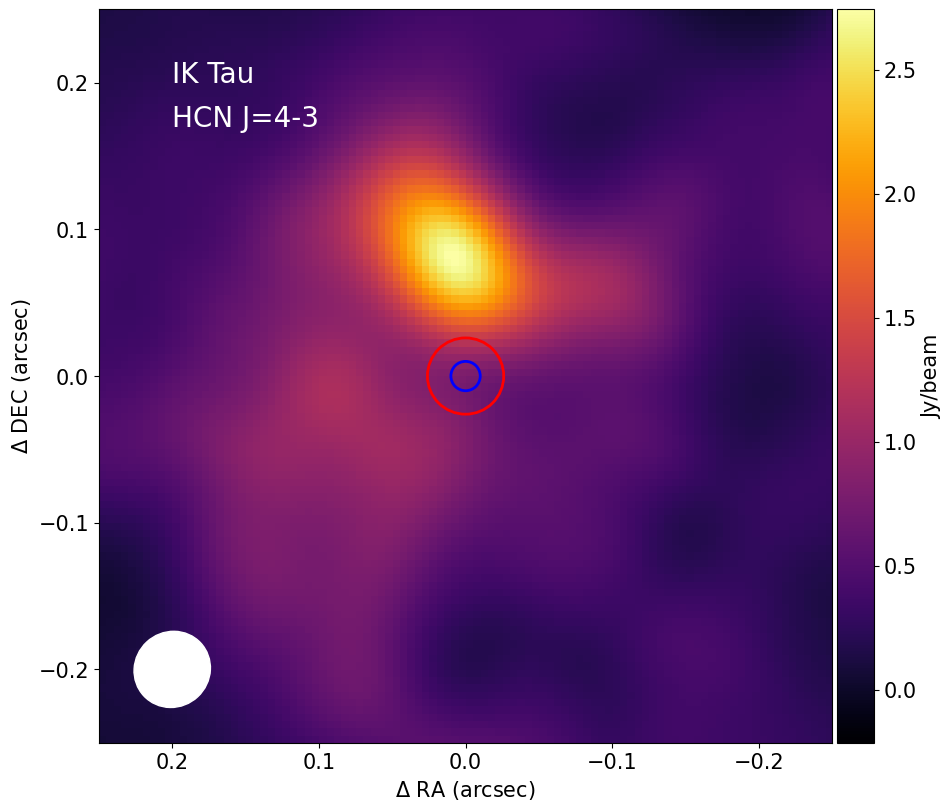}
\includegraphics[width=0.40\textwidth,clip]{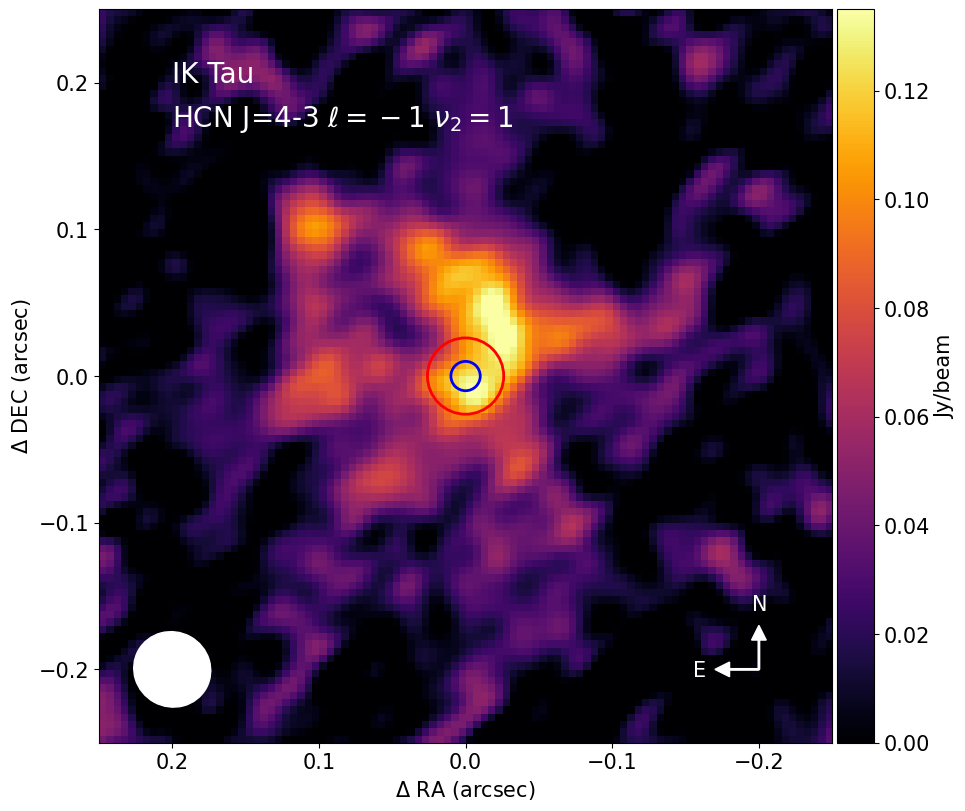}
\caption{Spatial distribution of the emission of HCN $J=4-3$ in the ground vibrational state (left) and $\nu_2=1$ state (right) integrated between $\pm 5$ km s$^{-1}$ centered on the $V_\textrm{sys}$, in the inner regions of R\,Crt (top) and IK\,Tau (bottom). The blue circle is the size of the AGB star \citep[as measured in the near infrared][]{Paladini2017, Richichi2005}, and the red circle is the size of the radio photosphere at millimeter wavelengths. The white ellipse in the bottom left corner represents the reconstructed beam.
}
\label{fig:emission}
\end{figure*}

This study focuses on two O-rich stars, R\,Crt and IK\,Tau, with low and intermediate mass loss rates, respectively (see Table\,\ref{tab:parameters}). The two sources were observed with the Atacama Large Millimeter/submillimeter Array (ALMA) in band 7 at high angular resolution for the continuum and molecular lines. The emission of the lines HCN $J$\,=\,4-3 in the ground vibrational state at 354.505473 GHz and HCN $J$\,=\,4-3 $\ell$\,=\,$-$1 in the $\nu_2=1$ vibrational state at 354.460436 GHz\footnote{Frequencies from the CDMS database \url{https://cdms.astro.uni-koeln.de}} were detected in both R\,Crt and IK\,Tau, and their spatial distribution is presented in Fig.\,\ref{fig:emission}. 

\begin{table}
\centering
\caption{Stellar and circumstellar parameters for R\,Crt and IK\,Tau.}
\label{tab:parameters}
\begin{tabular}{lcc}
\hline
\hline
Parameter & R\,Crt & IK\,Tau \\
\hline
RA (J2000) & 11$^h$00$^m$33$.^{\!\!s}$8075 & 03$^h$53$^m$28$.^{\!\!s}$9357  \\
DEC (J2000) & $-$19$^\circ$40$^\prime$30$\farcs$368 & +11$^\circ$24$^\prime$21$\farcs$334 \\
$V_{\mathrm{sys}}$ [km\,s$^{-1}$] & +11.5\,$^{(a)}$ & +34.5\,$^{(h)}$ \\
$P$ [day] & 160\,$^{(b)}$ & 470\,$^{(g)}$ \\
$D$ [pc] & 213$\pm$7\,$^{(c)}$ & 261$\pm$19\,$^{(g)}$ \\
$T_\star$ [K] & 2800\,$^{(d)}$ & 2234\,$^{(g)}$ \\
$L_\star$ [$L_\odot$] & 7000\,$^{(e)}$ & 7900\,$^{(g)}$ \\
$R_\star$ [mas] & 9.0\,$^{(f)}$ & 10.35\,$^{(h)}$ \\
$R_\star$ [cm] & 2.9\,$\times$\,10$^{13}$ & 4.0\,$\times$\,10$^{13}$ \\
$n_\star$ [cm$^{-3}$] & 1.27\,$\times$\,10$^{14}$\,$^{(e)}$ & 1.60\,$\times$\,10$^{14}$\,$^{(e)}$ \\
$r_c$ [cm] & 1.8\,$\times$\,10$^{14}$\,$^{(e)}$ & 2.0\,$\times$\,10$^{14}$\,$^{(g)}$ \\
$\tau_{10}$ & 0.05\,$^{(e)}$ & 1.0\,$^{(g)}$ \\
$T_d(r_c)$ [K] & 1000\,$^{(e)}$ & 1000\,$^{(g)}$ \\
$\dot{M}$ [$M_\odot$\,yr$^{-1}$] & 8.7\,$\times$\,10$^{-7}$\,$^{(e)}$ & 4.8\,$\times$\,10$^{-6}$\,$^{(g)}$ \\
$\delta$ & 0.65\,$^{(e)}$ & 0.65\,$^{(g)}$ \\
$v_\infty$ [km s$^{-1}$] & 10.0\,$^{(d)}$ & 17.5\,$^{(g)}$ \\
$\Psi$ & 890\,$^{(e)}$ & 130\,$^{(g)}$ \\
$R_{\mathrm{mm}}$ [mas] & 30\,$^{(e)}$ & 26\,$^{(e)}$ \\
$R_{\mathrm{mm}}$ [$R_\star]$ & 3.2 & 2.5 \\
\hline
\end{tabular}
\tablefoot{
$^{(a)}$\,\cite{Massalkhi2020},
$^{(b)}$\,GCVS \citep{GCVS}, 
$^{(c)}$\,Gaia DR3 \cite{gaiadr3}, 
$^{(d)}$\,\cite{Schoier2013},
$^{(e)}$\,this study, 
$^{(f)}$\,\cite{Khouri2020}, 
$^{(g)}$\,\cite{Massalkhi2024},
$^{(h)}$\,\cite{Adam2019}.
}
\end{table}

R\,Crt and IK\,Tau were observed with ALMA during Cycle 7 as part of project 2019.1.00801.S (P.I.: J. Champion). The observations were made with three different configurations (C-7, C-4, and C-2) with baselines ranging from 15 to 8282~m. The effective number of antennas ranged from 39 to 45 depending on their performance. The observations were taken around $\simeq 0.86$~mm (band 7) on 2019 Oct 6, 9, and 30, 2020 Jan 9, and 2021 Oct 31 and Nov 1. This study focuses on the data obtained with the most extended configuration. The observations provide maximum angular resolutions of $\simeq 0\farcs03$ and maximum recoverable scales (MRSs) of $\simeq 0\farcs6$ at 345~GHz for the most extended configuration. The mean precipitable water vapor ranges from 0.46 to 1.1~mm, which implies system temperatures of $100-200$~K for R\,Crt and  $125-250$~K for IK\,Tau.

The total spectral coverage was divided into four windows with widths of 1875~MHz approximately centered at 340.55, 342.50, 352.64, and 354.41~GHz rest frequencies. The spectral resolution was 1.129~MHz or $\simeq 1.0$~km~s$^{-1}$ at the observed frequencies, which is enough to spectrally resolve the lines from AGB envelopes. Flux, bandpass, and pointing were calibrated in the usual way with calibrators J1058+0133 for R\,Crt and J0423-0120 for IK\,Tau. J0407+0742 and J1048-1909 were periodically observed to calibrate phases (every $\simeq 1.4-1.7$~min for both targets in the extended configurations). J0401+0413 and J1035-2011 were observed five to six times during the runs as check sources. Using the measurements in the ALMA's Calibrator Catalogue, we estimate average amplitude uncertainties of $\simeq 7$\% and 4\% for R\,Crt and IK\,Tau, respectively. The visibility tables were recovered with the recommended CASA pipelines (versions 5.6.1-8 and 6.2.1.7; \citealt{mcmullin_2007}). These tables were subsequently exported to GILDAS\footnote{\url{https://www.iram.fr/IRAMFR/GILDAS}}, where mapping and data analysis were fully done. 

In this work, CLEANing was done using the Steer-Dewdney-Ito (SDI) method \citep{steer_1984}, which improves the mapping quality for extended emissions. Maximum angular resolutions of $0\farcs070\times 0\farcs029$ and $0\farcs047\times 0\farcs031$  were respectively reached for R\,Crt and IK\,Tau for the molecular line maps with uniform weighting. The restored maps have a Half-Power Beam Width (HPBW) of $0\farcs06$\,$\times$\,$0\farcs05$ and a Position Angle (P.A.) of 95$^\circ$ for R\,Crt, and a round Point Spread Function (PSF) with a HPBW of 0.05 arcsec$^{2}$ for IK\,Tau.

The use of the extended configuration lacking short baselines prevents us from properly describing structures larger than $\sim$\,$0\farcs6$. These structures are however not important here since we focus on the spatial distribution of HCN out to $\sim$\,$0\farcs2$ from the star. To evaluate how flux filtering could affect our results, we combined the data of the extended configuration with those from the more compact ones. From this exercise we concluded that the fraction of flux filtered out when using only extended configuration data below 20\,\%. In any case, and more importantly, flux filtering is relatively uniform across the different spatial locations within $\sim$\,$0\farcs2$ from the star, so that relative variations of the brightness, in which we are mostly interested here, remain unaffected.\\

The data for both targets were phase self-calibrated, taking advantage of the strong continuum fluxes. Two loops were enough to complete this task, resulting in an rms noise improvement in the continuum maps of a factor of $3.5-4.0$$^($\footnote{Phase self-calibration does not have a real impact on the molecular line maps with low rms noise improvements of up to 30\%.}$^)$ The signal-to-noise ratio S/R of the continuum emission for both sources is high ($\simeq 100-500$ depending on the weighting and the source) and the nominal positional accuracy can be taken as 10\% of the HPBW \citep{cortes2020alma}.

The continuum sources of R\,Crt and IK\,Tau are not resolved in the self-calibrated maps produced with uniform weighting at the detection level ($3\sigma$), where rms noises are 0.31 and 0.21~mJy~beam$^{-1}$ and intensity maxima are 30 and 80 mJy~beam$^{-1}$, respectively. Using natural weighting to reduce the noise level to 0.08 and 0.17~mJy~beam$^{-1}$ does not unveil any weak structure around the central component of the continuum.

From Gaussian-fitting to the non-self-calibrated continuum at 354\,GHz brightness distributions, we derive central positions of (RA, DEC) = (03$^h$53$^m$28$.^{\!\!s}$9357, +11$^\circ$24$^\prime$21$\farcs$334) and (11$^h$00$^m$33$.^{\!\!s}$8075, $-$19$^\circ$40$^\prime$30$\farcs$368) for the continuum sources of IK\,Tau and R\,Crt at millimeter wavelengths. Natural weighting was adopted this time to avoid noticeable structures present in the non-self-calibrated maps produced with uniform weighting that could affect the Gaussian fits. Asymptotic giant branch stars are known to have a radio photosphere originated by free-free radiation \citep{Reid1997}. The size of the radio photosphere was estimated at 354GHz with our ALMA data by fitting the emission above 3$\sigma$ level with a 2D Gaussian using the SciPy python library. Since we are interested in HCN lines around 354 GHz, the size of the radio photosphere was estimated from the ALMA continuum emission at 354 GHz. The derived sizes are given as $R_{\rm mm}$ in Table\,\ref{tab:parameters}, where $R_{\rm mm}$ corresponds to half the Full Width at Half Maximum (FWHM), where the free-free emission becomes optically thick.

\section{Model}
\label{sec:meth}

We are interested in the abundance distribution of HCN in R\,Crt and IK\,Tau, and therefore we need to convert the observed brightness distribution of the HCN lines into an abundance. It is therefore necessary to theoretically describe the excitation and solve the radiative transfer problem for HCN to properly model its emission around these stars. As mentioned in Sect.\,\ref{sec:obse}, the estimated amount of flux filtered in the observed maps in the region up to $\sim$\,$0\farcs2$ from the star is small, and in any case we base the comparison between the observation and model on the relative intensity variation as a function of the distance to the star, which is little affected by flux filtering.

We considered an AGB star surrounded by a spherical envelope of gas and dust. In Sect.\,\ref{sec:result} we discuss the impact of accounting for departures from spherical geometry, as indicated by the emission distribution of HCN in R\,Crt and IK\,Tau (see Fig.\,\ref{fig:emission}). The model can be described by a few input parameters (given in Table\,\ref{tab:parameters}). The star is characterized by an effective temperature $T_\star$, luminosity $L_\star$, and radius $R_\star$. We favor the stellar radii derived from near-infrared interferometric observations \citep{Khouri2020,Adam2019} over the values inferred from $T_\star$ and $L_\star$ using the Stefan-Boltzmann law. In the case of R\,Crt, the stellar radius derived from the adopted values of $T_\star$ and $L_\star$ is 2.5\,$\times$\,10$^{13}$ cm, which is somewhat smaller than the value of 2.9\,$\times$\,10$^{13}$ cm derived from near-infrared observations. For IK\,Tau, the stellar radius derived from $T_\star$ and $L_\star$ is 4.1\,$\times$\,10$^{13}$ cm, which is very close to the near-infrared radius of 4.0\,$\times$\,10$^{13}$ cm.

\begin{figure}
\centering
\includegraphics[width=0.53\textwidth,clip]{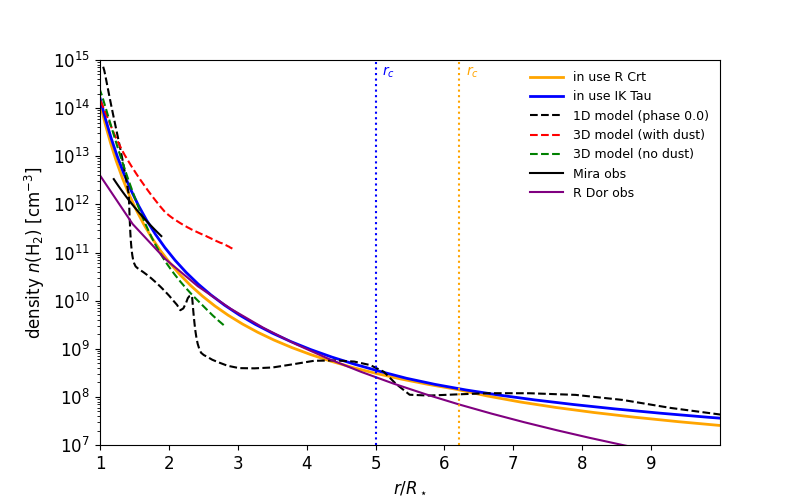}
\caption{Volume density of H$_2$ in the inner region of the envelopes around R\,Crt (solid yellow) and IK\,Tau (solid blue) as a function of the radial distance from the star compared with theoretical and observational density profiles. The vertical dotted lines indicate the condensation radii for each envelope. The 1D model is from \cite{Bladh2019} and the 3D model from \cite{Freytag2017}. The observational profiles for Mira and R\,Dor are from \cite{Vlemmings2019} and \citep{Khouri2024}, respectively.
}
\label{pressure}
\end{figure}

\begin{figure*}
\centering
\includegraphics[width=0.42\textwidth,clip]{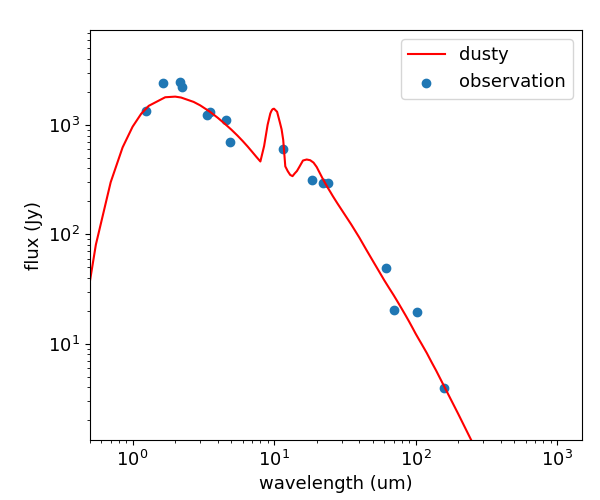}
\includegraphics[width=0.53\textwidth,clip]{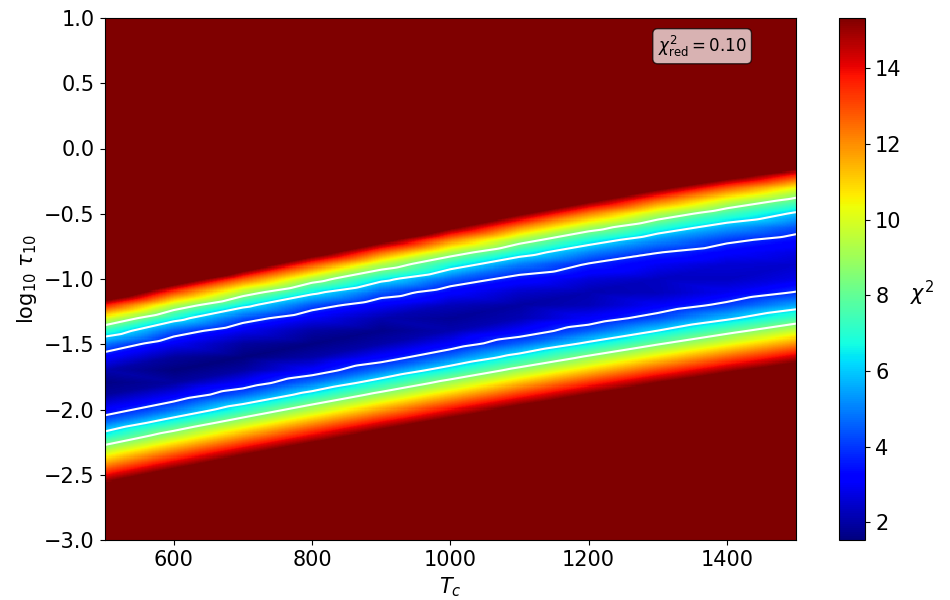}
\caption{Results from the SED analysis for R\,Crt. Left: SED calculated with the best-fit model compared with the observed fluxes (see Table\,\ref{tab:contfluxRCrt}). Right: $\chi^2$ as a function of the condensation temperature and the opacity at 10 $\mu$m. The $\chi^2_{red}$ value was calculated with 17 flux points. The white contours correspond to 1, 2, and 3$\sigma$ confidence levels (1$\sigma= 3.83$). } 
\label{paramdusty}
\end{figure*}

For the envelope, we assumed the presence of dust with a uniform gas-to-dust mass ratio, $\Psi$, only beyond a given radial distance, which can be rationalized as the dust condensation radius, $r_c$. The envelope is characterized by a mass loss rate, $\dot{M}$, and a terminal expansion velocity, $v_\infty$. The expansion velocity beyond $r_c$ is given as a function of the radial distance from the star, $r$, as
\begin{equation}
\varv_\textrm{exp}(r) = \varv_0 + (\varv_\infty - \varv_0)\left(1- \frac{r_c}{r} \right)^\beta,
\end{equation}
where we assume $\varv_0 = \varv_\infty/4$ and $\beta=1$ \citep{Decin2010}. For $r < r_c$ we assume $v_{\rm exp} = v_0$. 

The volume density of the gas, $n_g$, as a function of the radial distance is described by
\begin{equation}
n_g(r) =  \left\{
\begin{array}{ll}
n_\star \exp(-c \, (1-(R_\star/r)^b)) & \mbox{if } r<r_c \\
\dot{M} / (\overline{m}_g\,4 \pi \, \varv_\textrm{exp} \, r^2) & \mbox{if } r \geq r_c,
\end{array}
\right.
\end{equation}
where for $r$\,$>$\,$r_c$, we assume the usual mass conservation law. For $r$\,$<$\,$r_c$, we adopted a parametric expression that approximately describes the density fall off in the extended atmosphere of AGB stars according to hydrodynamic models \citep{Freytag2017,Bladh2019} and ALMA continuum observations \citep{Vlemmings2019,Khouri2024}. We adopted a photospheric pressure of 5\,$\times$\,10$^{-5}$ bar from the 3D model with dust from \cite{Freytag2017}. From this model, the density $n_\star$ was determined using the ideal gas law (given in Table \ref{tab:parameters}) with an exponent $b$ of 2, while the parameter $c$ was set to ensure continuity in the density before and after $r_c$. We note that densities in these very inner layers are subject to important uncertainties. The adopted radial density profiles for R\,Crt and IK\,Tau are shown in Fig.\,\ref{pressure}. The gas temperature as a function of radial distance is described by 
\begin{equation}
T_g(r) = T_\star \left( \frac{r}{R_\star}\right)^{-\delta},
\label{eq:temp}
\end{equation}
where the exponent $\delta$ is determined by fitting multiple CO lines (see below). 

\begin{figure*}
\centering
\includegraphics[width=0.51\textwidth,clip]{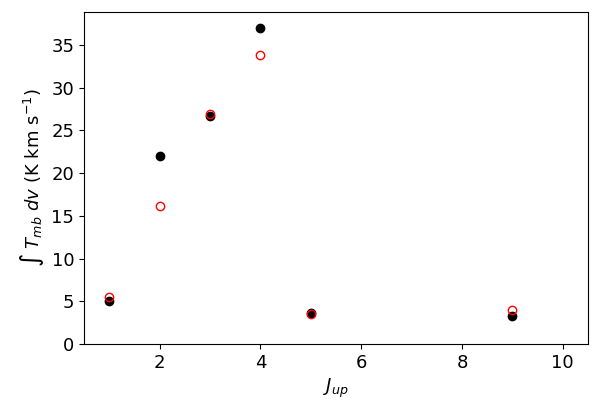}
\includegraphics[width=0.45\textwidth,clip]{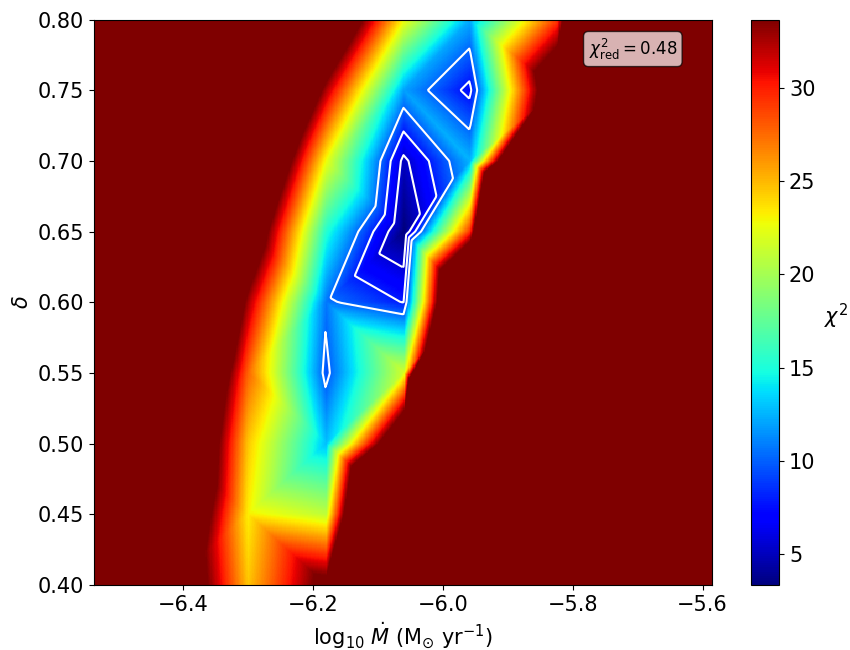}
\caption{Results from the CO emission analysis for R\,Crt. Left: Line intensities calculated with the best-fit model compared with the observed values (see Table\,\ref{tab:co_transitions}). Right: $\chi^2$ as a function of the mass loss rate and the exponent of the gas temperature radial profile. The $\chi^2_{red}$ value was calculated with six flux points; the white contours correspond to the 1, 2, and 3$\sigma$ confidence levels (1$\sigma = 5.66$).} 
\label{paramCO}
\end{figure*}

The parameters entering in the above physical description of R\,Crt and IK\,Tau envelopes are given in Table\,\ref{tab:parameters}. In the case of IK\,Tau, the parameters were taken from \cite{Massalkhi2024}, while for R\,Crt we performed an analysis similar to that by these authors. Briefly, for R\,Crt we adopted the distance \citep{gaiadr3}, stellar effective temperature \citep{Schoier2013}, and terminal expansion velocity \citep{Massalkhi2020} from the literature and derived the dust condensation radius, luminosity, gas-to-dust mass ratio, mass loss rate, and gas and dust temperature profiles. To describe the dusty envelope, we used the code DUSTY\,V2\footnote{\url{http://faculty.washington.edu/ivezic/dusty_web/}} \citep{dusty}, which solves the continuum radiative transfer for a spherical dusty envelope, including dust scattering, absorption, and emission. We considered spherical grains with a radius of 0.1 $\mu$m, with the optical properties of warm silicate dust from \cite{Suh1999}. A grid of models was run, where the optical depth at 10 $\mu$m, $\tau_{10}$, and the dust temperature at the condensation radius, $T_c$, were varied. Moreover, a $\chi^2_\textrm{red}$ analysis was performed to find the model that best reproduces the observed spectral energy distribution (SED). Photometric fluxes for R\,Crt were taken from the Vizier database\footnote{\url{http://vizier.cds.unistra.fr/vizier/sed/}} and are presented in Table\,\ref{tab:contfluxRCrt}. We adopted $20 \%$ uniform uncertainties for all measurements. The best-fit model is compared with the observed fluxes in the left panel of Fig.\,\ref{paramdusty}, while the right panel shows the $\chi^2_\textrm{red}$ parameter as a function of $T_c$ and $\tau_{10}$. The output of the best-fit DUSTY model provides the optimal values of $T_c$, $\tau_{10}$, $r_c$, the dust mass loss rate (from which we can estimate the gas-to-dust mass ratio), and the dust temperature radial profile. \\

To describe the gaseous envelope of R\,Crt, we proceeded as in \cite{Massalkhi2024} and carried out a grid of non local thermodynamic equilibrium (LTE) radiative transfer models for CO, varying the mass loss rate and exponent $\delta$ of the gas temperature radial profile. The fractional abundance of CO with respect to H$_2$ at the stellar photosphere was set to $2.0 \times 10^{-4}$ relative to H$_2$, a value typical of O-rich envelopes \citep{Kahane1994}\footnote{Additional models were run as a consistency check, varying the assumed CO abundance by a factor of ~2. The resulting best-fit parameter ranges remain sufficiently narrow to support the adopted value for the CO abundance.}. We included CO in the vibrational states $v=0$ and $v=1$ to account for infrared pumping and used the Monte Carlo algorithm of RATRAN \citep{ratran}, modified to enable a layer-dependent number of levels, to solve the statistical equilibrium (see \citealt{Massalkhi2024}). A $\chi^2_\textrm{red}$ analysis was carried out to find the best-fit model by comparing the calculated velocity-integrated intensities with the observed intensities of multiple CO lines (see Table\,\ref{tab:co_transitions}). We adopted $20 \%$ uniform uncertainties for all CO lines. We considered the HPBW of the different telescopes during the fitting process. The $\chi^2_\textrm{red}$ parameter is shown as a function of the mass loss rate and $\delta$ in the right panel of Fig.\,\ref{paramCO}, while the left panel compares the best-fit CO line intensities with the observed ones. The fit to CO is good, and the mass loss rate and exponent of the gas temperature law are well determined. The derived mass loss rate and exponent of the gas temperature for R\,Crt are 8.7\,$\times$\,10$^{-7}$ and 0.65, respectively.

Once the model of the dusty and gaseous envelope was properly described for R\,Crt and IK\,Tau, we carried out excitation and radiative transfer calculations to model the observed emission of HCN and constrain its abundance distribution. Unlike in the previous case of CO, where we focused on the distribution at large scales, here we are interested in the HCN distribution in the inner circumstellar layers. We considered that HCN forms at some initial radial distance $r_h$ from the star, so that there is a hole or void of HCN in the central region around the star. Our main aim is to constrain this initial radius from the ALMA observations and HCN model. The abundance distribution beyond the initial radius was assumed to be given by
\begin{equation}
f(r) = f_0 \exp \Bigg[ - \bigg( \frac{r}{r_e} \bigg)^2 \Bigg],
\end{equation}
where the HCN abundance at the initial radius is $f_0$, and the $e$-folding radius, $r_e$. Following \cite{Schoier2013}, the abundance distribution at the initial radius for R\,Crt and IK\,Tau are 3.5\,$\times$\,10$^{-7}$ and 4.3\,$\times$\,10$^{-7}$. The $r_e$ value is evaluated through the empirical expression in Eq.\,(3) of \cite{Schoier2013} and are 1.0\,$\times$\,10$^{16}$ cm and 2.0\,$\times$\,10$^{16}$ cm for R\,Crt and IK\,Tau, respectively.

We considered the existence of a radiophotosphere around the star where opacity is driven by free-free emission. The optical depth of the ionized gas between $R_\star$ and $R_{\rm mm}$ was evaluated as \citep{Mezger1967}
\begin{equation}
\tau = 8.235 \times 10^{-2} ~ \nu^{-2.1} ~ T_e^{-1.35} ~ n_e^2 ~ \Delta s,
\end{equation}
where $T_e$ is the electron temperature in kelvin (approximated as the gas temperature at the stellar surface), $n_e$ is the electron volume density per cubic centimeters (we assumed an ionization degree of 10$^{-5}$ based on chemical equilibrium calculations; \citealt{Agundez2020}), $\nu$ is the frequency in gigahertz, and $\Delta s$ is the path length in parsec.

The excitation and radiative transfer of HCN was solved in non-LTE adopting the large velocity gradient (LVG) formalism. The code has been used previously in, for example, \cite{Agundez2012} and \cite{Massalkhi2024}. We considered the first 41 rotational levels within the ground vibrational state and within the vibrationally excited states $v_2$\,=\,1, $v_2$\,=\,2, $v_3$\,=\,1, and $v_1$\,=\,1, where $v_2$ refers to the bending mode lying at 713.5 cm$^{-1}$, $v_3$ to the CN stretching at 2096.8 cm$^{-1}$, and $v_1$ stands for the CH stretching mode at 3311.5 cm$^{-1}$. The energy levels and radiative transition Einstein coefficients were taken from the HITRAN2020 database \citep{hitran}. The collisional downward rate coefficients for pure rotational transitions were taken from \cite{Hernandez-Vera2017} for collisions with a para/ortho H$_2$ (adopting an ortho-to-para ratio of 3) and from \cite{Dumouchel2010} for collisions with He, adopting a solar abundance of 0.17 relative to H$_2$ for He \citep{Asplund2009}. For ro-vibrational transitions, we adopted the collision rate coefficients of the analogous pure rotational transition but diminished by a factor of 10,000 \citep[typical value for ro-vibrational excitation of SiO and CS][]{Balanca2017, Lique2007}. Finally, all upward excitation rates were computed systematically from the downward ones applying detailed balance.

\section{Results}
\label{sec:result}

Chemical equilibrium calculations predict low abundances for HCN in the innermost shells of O-rich envelopes \citep[$< 10^{-10}$ relative to H$_2$;][]{Agundez2020}. Observations on the other hand find that HCN is present in the outer envelope with abundances relative to H$_2$ of $\sim$\,10$^{-7}$ \citep{Schoier2013}. While HCN is observed in many highly excited vibrational states in C-rich objects such as IRC\,+10216 \citep{Cernicharo2011,Cernicharo2013, Velilla-Prieto2023}, with some showing maser emission \citep{Menten2018, Jeste2022, Yang2025}, the same is not true for O-rich stars \citep{Ohnaka2025}. This suggests that the bulk of HCN must form in the inner regions around O-rich stars but not at the stellar surface, which should manifest as an inner void or “hole” in the HCN abundance distribution.

To constrain the formation radius of HCN in R\,Crt and IK\,Tau, we compared the calculated brightness radial profile and the observed azimuthal average of the brightness distribution. We focused on the velocity range $\pm$\,5 km s$^{-1}$ around the systemic velocity, which probes the material in the inner accelerating region of the envelope. An expansion velocity close to the terminal value, 10.0 km s$^{-1}$ in R\,Crt and 17.5 km s$^{-1}$ in IK\,Tau, is already reached at $10-20$ $R_\star$ \citep{Decin2010, Decin2018}. By considering velocities of $\pm$\,5 km s$^{-1}$, we therefore excluded the contribution from distant material along the line of sight and isolated the HCN emission formed in the innermost regions. We note that thermal and turbulent velocities in these inner regions are expected to be smaller than 5 km s$^{-1}$ \citep{Agundez2012}. We tested several sizes of holes in the radial abundance profile of HCN to constrain its formation radius, $r_h$. We explored models with $r_h$ ranging from 1~$R_\star$ to 10~$R_\star$, where $r_h$\,=\,1~$R_\star$  corresponds to a reference model where HCN forms at the photosphere. We note that the observed brightness distribution of HCN is not symmetric around the star in either IK\,Tau or R\,Crt. As shown in the HCN $J$\,=\,4-3 maps of Fig.\,\ref{fig:emission}, there is a bright spot in R\,Crt in the northwest and in IK\,Tau in the north. This can be due to the asymmetrical ejection of matter at the stellar surface \citep{Velilla-Prieto2023} or to the presence of a companion around the AGB star \citep{Decin2020}, both of which could create asymmetric structures around the star. Here, we used a simple observed azimuthal average in which we averaged regions including and excluding the corresponding bright spots. Nevertheless, our conclusions on the formation radius of HCN are significantly affected by the bright clump, as discussed in detail in Appendix\,\ref{app:angle}, where we consider specific regions of the maps, with and without the bright clump.

\begin{figure}
\centering
\includegraphics[width=0.45\textwidth,clip]{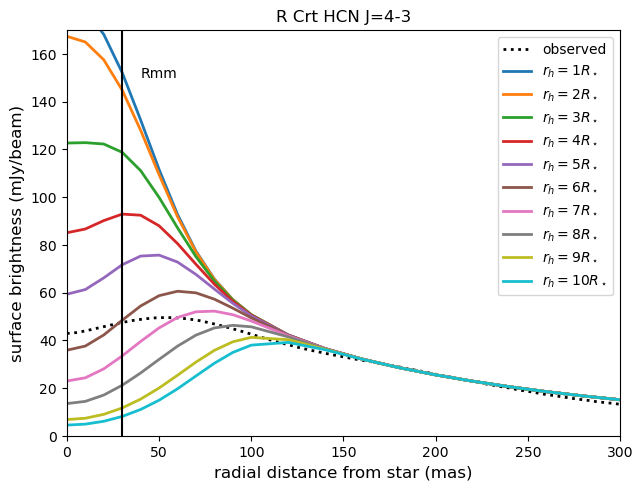}
\includegraphics[width=0.45\textwidth,clip]{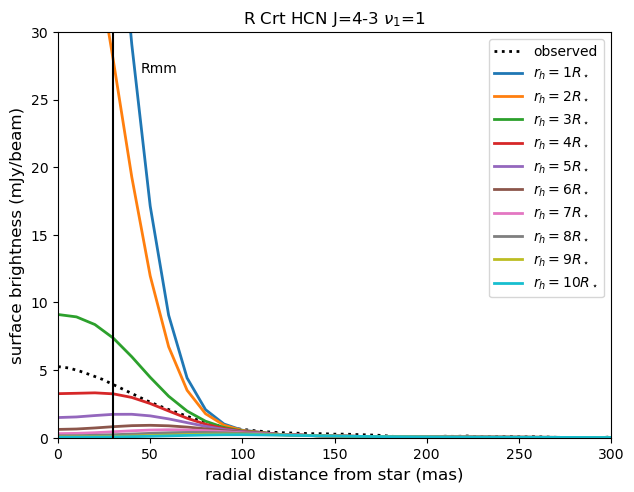}
\caption{Azimuthally averaged HCN $J$ = 4–3 line emission around the systemic velocity in the envelope of R\,Crt. Top: Emission of the $J$\,=\,4-3 ground vibrational state line. Bottom: $J$\,=\,$4-3$ in the $\nu_2 = 1$ vibrational state. The dotted black curve corresponds to the observation, while the solid curves represent the models with $r_h$ between 1 and 10 $R_\star$. In both cases the line emission is integrated over $\pm$\,5 km s$^{-1}$ around the systemic velocity. The vertical line $R_\textrm{mm}$ represents the radiophotosphere radius. 
}
\label{HCN-RCrt}
\end{figure}

\begin{figure}
\centering
\includegraphics[width=0.45\textwidth,clip]{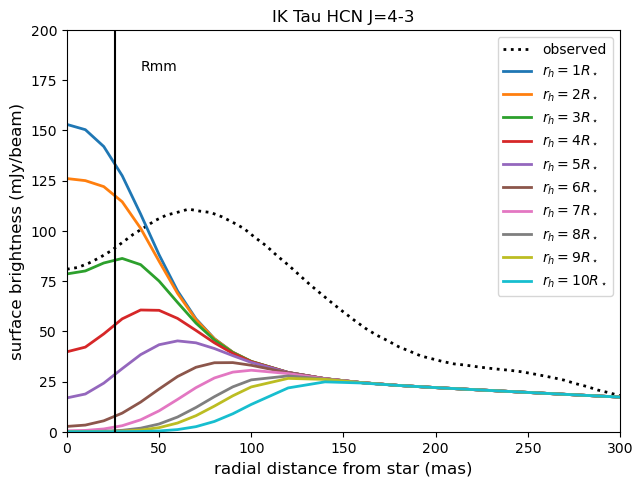}
\includegraphics[width=0.45\textwidth,clip]{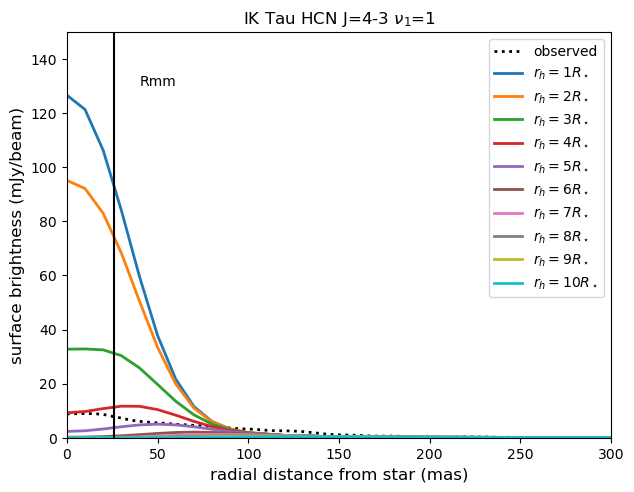}
\caption{Same as Fig.\,\ref{HCN-RCrt} but for IK\,Tau.}
\label{HCN-IKTau}
\end{figure}

The upper panel in Fig.\,\ref{HCN-RCrt} compares the calculated and observed brightness radial profiles of the HCN $J$\,=\,4–3 line in the ground vibrational state for R\,Crt, while the upper panel in Fig.\,\ref{HCN-IKTau} presents the same for IK\,Tau. These plots make comparisons between the models (solid lines) and observations (dotted lines), both integrated over the same velocity range, $\pm$\,5 km s$^{-1}$. We first note that for both R\,Crt and IK\,Tau, the model with $r_h$\,=\,1 $R_\star$ fails to reproduce the observed brightness distribution. Therefore, a model in which HCN is present down to the stellar surface can be excluded, implying that HCN must form at a certain distance from the star.

For R\,Crt, if we focus on the $J$\,=\,4-3 line and examine the morphology of the radial brightness profile, the best match is achieved for $r_h$\,=\,4-6 $R_\star$. If we examine the radius at its intensity maximum, the best match is achieved at 66 mas in agreement with observations, while the models with the closest intensity maxima are those with $r_h$\,=\,5 $R_\star$. Finally, if we evaluate the agreement between the model and observation through the $\chi^2$ parameter, the best-fit model is $r_h$\,=\,6 $R_\star$. In this case we computed $\chi^2$ using 20 points with radii between 0 and 300 mas and obtain reduced $\chi^2$ values of 0.0774, 0.0652, 0.0326, 0.0131, 0.0043, 0.0007, 0.0015, 0.0042, 0.0083, and 0.0106 for the models with $r_h$\,=\,1 to 10 $R_\star$, respectively. If we examine the $J$\,=\,4-3 $\nu_2$ line, the model with the lowest $\chi^2_\textrm{red}$ is that with $r_h$\,=\,4 $R_\star$.

In the case of IK\,Tau, a comparison between the observed and calculated azimuthal averages of the $J$\,=\,4-3 line indicates that models with $r_h$\,=\,3-5 $R_\star$ provide the best overall agreement. The model with $r_h$\,=\,5 $R_\star$ produces an intensity maximum closest to the observed value (55 mas), while the model with the lowest $\chi^2_\textrm{red}$ corresponds to $r_h$\,=\,3 $R_\star$ (the $\chi^2_\textrm{red}$ values are 0.037, 0.029, 0.025, 0.035, 0.052, 0.075, 0.087, 0.097, 0.101, and 0.109 for the models with $r_h$\,=\,1, to 10 $R_\star$, respectively). In the case of the vibrationally excited line, the model with the lowest $\chi^2_\textrm{red}$ correspond to $r_h$\,=\,4 $R_\star$. 

After using different criteria to define the agreement between the observation and model, we conclude that the formation radius of HCN in R\,Crt is in the 5-6 $R_\star$ range. In the case of IK\,Tau, we find that HCN must be formed in a region located at 3-5 $R_\star$. Therefore, the formation radius of HCN is found to be very similar in these two O-rich AGB envelopes, despite having a different mass loss rate. \\

The observations of the HCN $J$\,=\,4-3 line show a bright spot in both R\,Crt and IK\,Tau that could impact the results. This raises the question of whether the global azimuthal average retrieved from the observations is representative of the global structure or whether the profile only shows the effect of the bump. We provide an analysis in Appendix\,\ref{app:angle} of the radial structure for two different regions: one with and another without the bump. We confirm that the conclusions regarding the formation radius of HCN remain similar regardless of whether or not we focus on the directions intersecting the bright spot.

\section{Discussion}
\label{sec:disc}

Although several molecules are known to be anomalously abundant in the inner envelopes of AGB stars, most are extremely challenging to constrain observationally, which limits our ability to accurately determine their formation radii and the dominant nonequilibrium processes at play. HCN, by contrast, is one of the most favorable tracers for such an investigation because it presents relatively bright lines observable with ALMA at high angular resolutions. CS is another molecule anomalously abundant in O-rich stars, but its emission is typically lower than that of HCN. Previous ALMA studies of CS in O-rich envelopes report relatively low line intensities \citep[e.g.,][]{Danilovich2019}. In that study, the authors did not consider a hole but a depletion of abundance in the inner region of IK\,Tau and W\,Hya. However, the modest angular resolution of their observations (100-150 mas) prevents an accurate determination of the formation radius of CS. 
Other species are even more difficult to access. NH$_3$ is also anomalously abundant, but it cannot be observed with ALMA because its fundamental transition at 572\,GHz is inaccessible from the ground. It can only be observed in the infrared through ro-vibrational lines and at centimeter wavelengths through the inversion lines. In C-rich stars, water is also problematic because ground-based observations are highly affected by atmospheric opacity for lines in the ground vibrational state, while lines from excited vibrational states are expected to be weak. PH$_3$ is expected to be faint in the millimeter domain. Its tentative identification in ALMA data for IRC\,+10216 with an angular resolution of $0\farcs6$\,$\times$$0\farcs5$ by \cite{Manna2024} was later shown to be incorrect \citep{Agundez2025}. This suggests that PH$_3$ is probably extended and difficult to target in the very inner regions, where its chemical origin could be constrained. Finally, SiH$_4$ is nonpolar and therefore invisible in the (sub)millimeter domain. 

HCN has been mapped around several O-rich AGB stars with a moderate angular resolution. For example, W\,Hya was mapped with an angular resolution of $0\farcs55$\,$\times$\,$0\farcs40$ \citep{Muller2008}, while IK\,Tau and R\,Dor were mapped with a resolution of 150 mas. These observations showed that HCN appears concentrated around the star, but the modest angular resolution and the lack of a radiative transfer model prevented researchers from precisely constraining the formation radius of HCN in these objects. The ALMA observations of R\,Crt and IK\,Tau presented here allowed us to constrain the formation radius of HCN. Based on our comparison between the observed and modeled azimuthally averaged radial profiles of the $J$\,=\,4-3 lines in the ground and $\nu_2$\,=\,1 vibrational states, the formation radius of HCN is constrained to 4-6 $R_\star$ and 3-5 $R_\star$ in R\,Crt and IK\,Tau, respectively. This void in the abundance of HCN in the central regions can be explained by shock-driven nonequilibrium chemical processes caused either by stellar pulsation and convective motions or by UV radiation. 

Models including shocks induced by stellar pulsation find that in O-rich stars HCN is formed in the inner wind, between 1 and 2 $R_\star$, and acts as a parent species throughout the envelope \citep{Duari1999, Marigo2016}. The models built by \cite{Cherchneff2006} for TX\,Cam find that HCN readily forms at 1 $R_\star$, while the more recent model by \cite{Gobrecht2016} built for IK\,Tau predicts that HCN appears at approximately 4 $R_\star$, which is consistent our observations here. Recently, \cite{Ohnaka2025} observed HCN at high angular resolutions (17-20 mas) in the low mass-loss-rate AGB star W\,Hya ($\sim$\,$10^{-7}$ $M_\odot$\,yr$^{-1}$; \citealt{Khouri2014}) and detected vibrationally excited HCN close to the star, down to $\sim$\,1.4\,$R_\star$. However, that study did not attempt to use a radiative transfer model to precisely locate the formation radius of HCN. \cite{Ohnaka2025} concluded that the the observation of HCN so close to the AGB star is consistent with shock-induced chemistry, although it may be too soon to conclude that this is the true nonequilibrium process responsible for its origin. 

Another possibility of nonequilibrium phenomenon involves UV photons of different origins. Models that consider photochemistry due to interstellar UV radiation penetrating from outside through a clumpy envelope predict that HCN forms at radial distances beyond 4 $R_\star$, with the exact location depending on the assumptions about the clumpiness of the envelope \citep{Agundez2010,VandeSande2018}. This scenario would be consistent with our findings only in the case of a high degree of clumpiness in the CSE. The origin of the UV field could also be internal. Ultraviolet radiation has been detected around some AGB stars \citep{Sahai2008,Montez2017}. Such UV fields can arise from binary companions or accretion disks \citep{Sahai2018}. It has been shown that 80\,$\%$ of AGB stars might have a companion \citep{Decin2020}, and the spiral structure of the CSEs can be explained by the presence of a companion \citep{Decin2015,Quintana2017,ElMellah2020}. The distribution of NaCl in IK\,Tau shows a spiral-like shape \citep{Coenegrachts2023}, which could be consistent with a potential companion, although it is not possible to identify the type of companion and whether or not it could be an important UV source. Ultraviolet radiation might also come from the chromospheric activity of the AGB star \citep{Montez2017}. The presence of internal UV radiation can impact the composition of the internal circumstellar regions via photochemistry \citep{VandeSande2019,VandeSande2022}. 
Only putting constraints on the strength of the UV field and the degree of porosity could provide more precise predictions for the abundance and formation radius. Internal X-rays could also drive important departures from chemical equilibrium. \cite{Alonso2025} have shown that molecules such as HCO$^+$ and HNC can be significantly enhanced in the inner regions of O-rich envelopes due to the action of X-rays, although HCN is not particularly enhanced via this mechanism.

\section{Conclusions}
\label{sec:Conclusions}

We presented ALMA high angular resolution observations of HCN in the O-rich AGB stars R\,Crt and IK\,Tau and carried out non-LTE radiative transfer calculations to derive the spatial distribution of HCN in the inner circumstellar regions. We constrain the formation radius of HCN to 4-6 $R_\star$ for R\,Crt and 3-5 $R_\star$ for IK\,Tau. These findings rule out efficient formation at the stellar photosphere but also beyond 5 $R_\star$. The derived formation radii are consistent with some models involving shock-induced chemistry but also with some scenarios where photochemistry driven by UV photons, either internal or external, is the main disequilibrium process. Detailed models tailored to specific objects would allow us to better distinguish between these two disequilibrium origins of the anomalously abundant molecule HCN in O-rich envelopes.

\begin{acknowledgements}

This work has been funded by Spanish Ministerio de Ciencia, Innovación, y Universidades (MICIU), Agencia Estatal de Investigaci\'on (AEI/10.13039/501100011033), and European Union (ESF+) through grants PID2020-117034RJ-I00 (MICIU, AEI), PID2023-147545NB-I00 (MICIU, AEI), and RYC2023-045648-I (MICIU, AEI, ESF+). Models have been run using the computational resources provided by the DRAGO computer cluster managed by SGAI-CSIC and the Galician Supercomputing Center (CESGA).

\end{acknowledgements}

\bibliographystyle{aa}
\bibliography{ref}

\begin{appendix}

\section{Observational Data for R\,Crt}

\begin{table}[ht!]
\centering
\caption{Observed photometric fluxes of R\,Crt extracted from the Vizier database (\url{http://vizier.cds.unistra.fr/vizier/sed/}).}
\label{tab:contfluxRCrt}
\begin{tabular}{cccc}
\hline
\hline
Wavelength ($\mu$m) & Flux (Jy) & Catalog \\
\hline
1.24 & 1340   & 2MASS \\
1.65 & 2400   & 2MASS \\
2.16 & 2460   & 2MASS \\
2.22 & 2220   & DIRBE \\
3.35 & 1220   & unWISE \\
3.52 & 1320   & DIRBE \\
4.60 & 1100   & unWISE \\
4.89 & 694    & DIRBE \\
11.6 & 601    & IRAS \\
18.4 & 314    & AKARI \\
22.1 & 295    & WISE \\
23.9 & 295    & IRAS \\
61.8 & 49.5   & IRAS \\
70   & 20.4   & Herschel/PACS \\
102  & 19.6   & IRAS \\
160  & 3.95   & Herschel/PACS \\
\hline
\end{tabular}
\end{table}

\begin{table}[ht!]
\caption{Observed velocity-integrated intensities of CO in R\,Crt.}
\label{tab:co_transitions}
\centering
\begin{tabular}{ccccc}
\hline
\hline
$J_u$-$J_l$ & $\nu$ & Telescope & $\int T_{\rm mb}\,dv$ & Reference \\
 & (GHz) & Telescope & (K km/s) & \\
\hline
1-0 & 115.271  & SEST  & 5.0   & 1 \\
2-1 & 230.538  & APEX  & 22.0  & 2 \\
3-2 & 345.796  & APEX  & 26.7  & 2 \\
4-3 & 461.040  & APEX  & 37.0  & 2 \\
5-4 & 576.268  & HIFI  & 3.57  & 3 \\
9-8 & 1036.912 & HIFI  & 3.26  & 3 \\
\hline
\end{tabular}
\tablefoot{
(1) \cite{Kerschbaum1999}. (2) \cite{Bergman2020}, from APEX Pointing Catalogue. (3) \cite{Danilovich2015}.
}
\end{table}

\section{Azimuthal profiles depending on the angles}
\label{app:angle} 

\begin{figure*}[hb!]
\centering
\includegraphics[width=0.45\textwidth,clip]{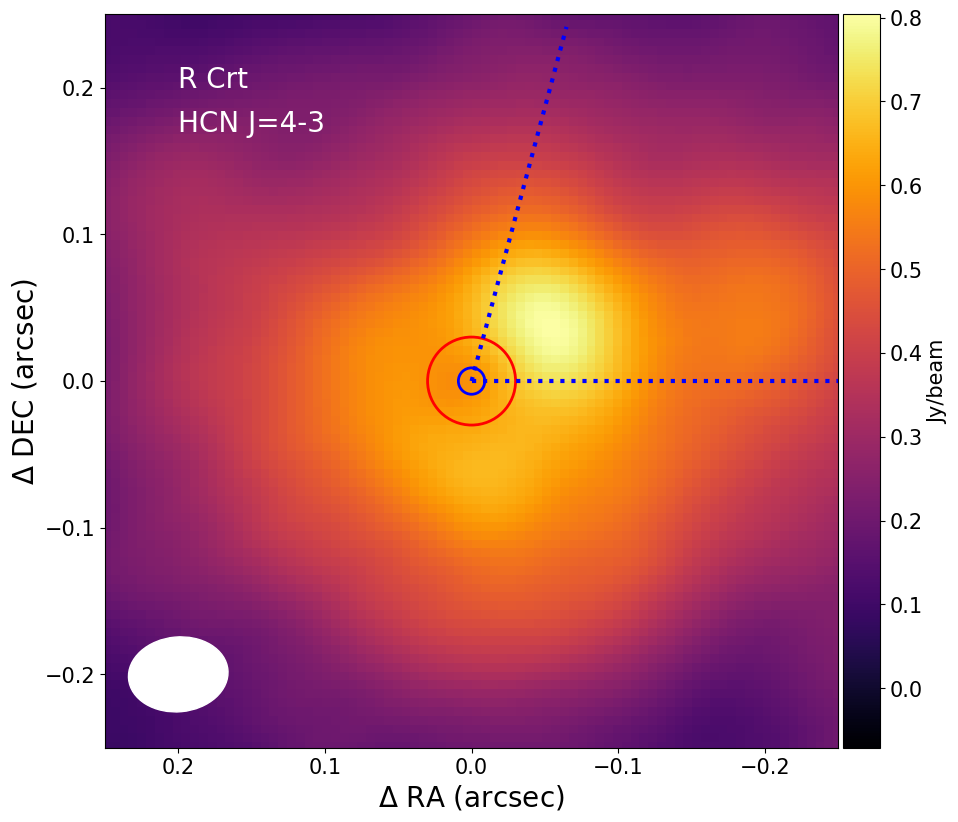}
\includegraphics[width=0.45\textwidth,clip]{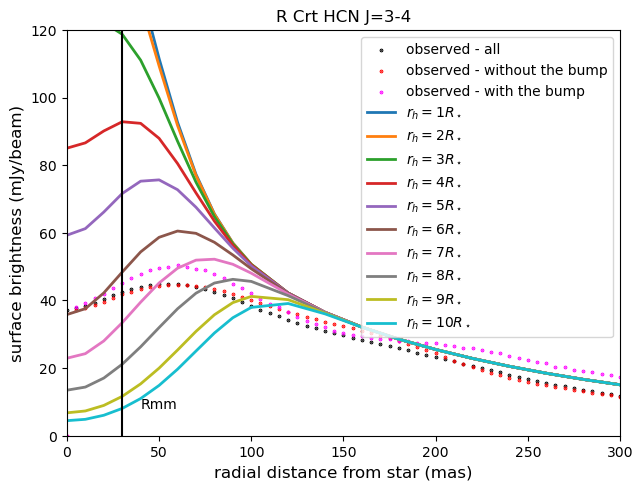}
\caption{Left: Spatial distribution of the emission of HCN $J$\,=\,4-3 in R\,Crt. The dotted blue lines represent the limits of the two regions studied, with and without the bright spot. Right: Azimuthal averaged HCN $J$\,=\,4-3 line emission around the systemic velocity $\pm$\,5 km s$^{-1}$. The dotted curves correspond to the ALMA observations (black one is the same presented in Fig.\,\ref{HCN-RCrt}) and the solid ones to the models with $r_h$ from 1 to 10 $R_\star$. The vertical line indicates the radiophotosphere radius, $R_{\rm mm}$.}
\label{HCN-bump-RCrt}
\end{figure*}

\begin{figure*}
 \centering
 \includegraphics[width=0.45\textwidth,clip]{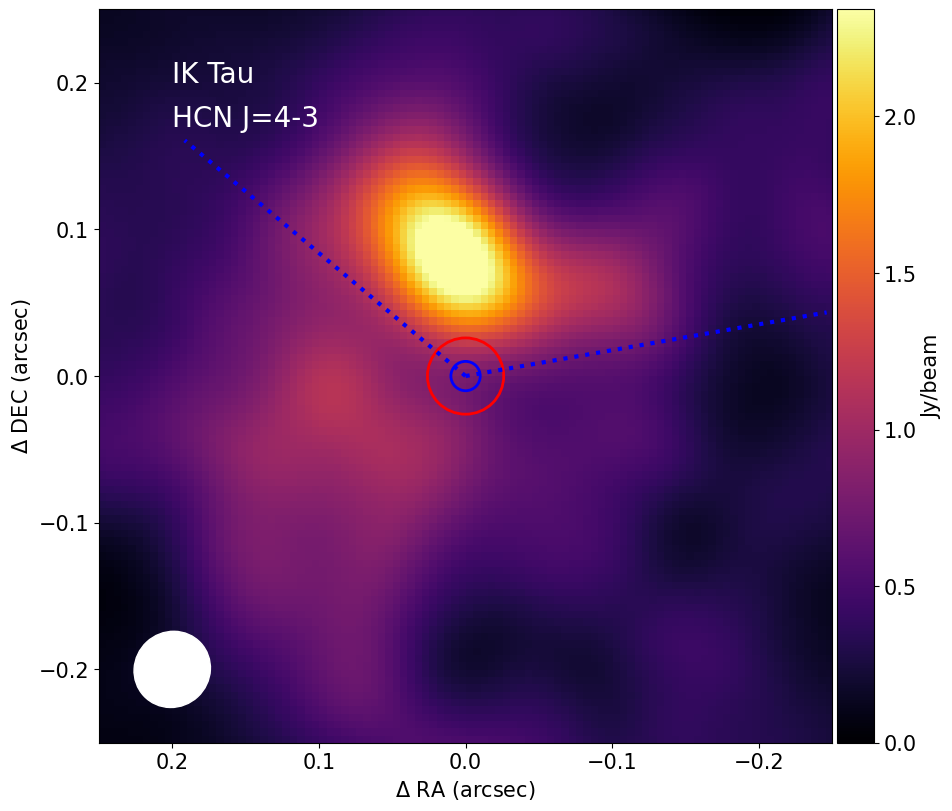}
 \includegraphics[width=0.45\textwidth,clip]{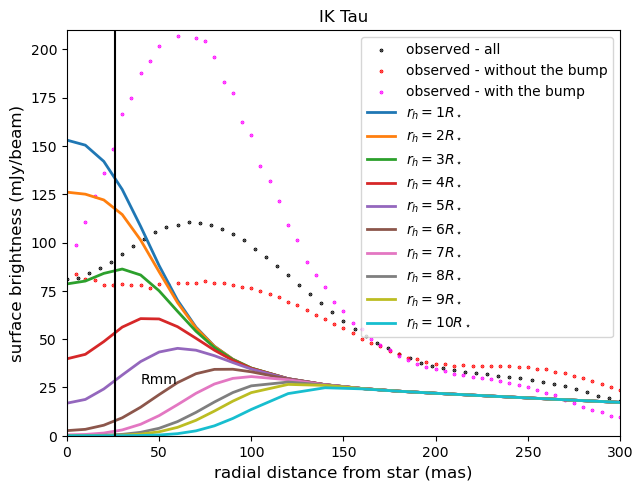}
  \caption{Same is Fig.\,\ref{HCN-bump-RCrt} for IK\,Tau.}
  \label{HCN-bump-IKTau}
\end{figure*}

In the HCN $J$\,=\,4-3 maps of R\,Crt and IK\,Tau there is a dominant bright clump in a specific direction (northwest in R\,Crt and north in IK\,Tau). The shape of the azimuthal averages derived in Sect.\ref{sec:result} could be completely dominated by this bright spot. One may wonder whether the azimuthal averaged profile is representative of the global structure. To shed light on this point and see how robust are our conclusions on the formation radius of HCN, we extracted radial strips at different position angles to study only the region with the bright spot and without it. The dotted blue lines in the left panels of Fig.\,\ref{HCN-bump-RCrt} for R\,Crt and Fig.\,\ref{HCN-bump-IKTau} for IK\,Tau delimitate these two regions (with and without bump). 

We consider three cases in which the observed brightness is azimuthally averaged over three different regions: (1) the full image, (2) the region without the bump, and (3) the region with the bump. In R\,Crt the bump is located between P.A. 270 and 345$^\circ$\footnote{P.A. is defined from north to the east (to the left on the images).} (see left panel in Fig.\,\ref{HCN-bump-RCrt}), while in IK\,Tau the bump is located between P.A. 280 and 50$^\circ$ (see left panel in Fig.\,\ref{HCN-bump-IKTau}). 

In R\,Crt, the observed azimuthal average of cases (1), (2), and (3) are very similar, with case (3) being slightly more intense (see right panel in Fig.\,\ref{HCN-bump-RCrt}). A similar analysis to that carried out for case (1) in Sect.\,\ref{sec:result} yields $r_h$\,=\,6 $R_\star$ for both cases (2) and (3). This is in line with the range of $r_h$ values of 4-6 $R_\star$ derived in case (1) and presented in Sect.\,\ref{sec:result}. In the case of IK\,Tau, the azimuthal averages of cases (1), (2), and (3) are noticeably different (see right panel in Fig.\,\ref{HCN-bump-IKTau}). The profile of case (2) is smoother than that of (1), although the flux at the central position is higher than at any other larger radius. This occurs because the radial profile has been computed by convolving the brightness distribution with a PSF equal to the beam size, and at the central position there is some contamination from the bright spot. The profile from case (3) is the most abrupt one, catching the marked contrast between the bright spot and the rest of regions. We derive $r_h$\,=\,4-6 $R_\star$ for case (2) and $r_h$\,=\,3-5 $R_\star$ for case (3), to be compared with the range $r_h$\,=\,3-5 $R_\star$ derived in case (1) and presented in Sect.\,\ref{sec:result}. In IK\,Tau, excluding the region with the bump, case (2), leads to a slight increase of $r_h$ by just one stellar radius. 

In summary, the ranges of $r_h$ derived when the bump is excluded from the analysis or when only the bright spot regions are considered, cases (2) and (3), respectively, are similar to that obtained when all azimuthal directions are averaged, case (1). We therefore consider that the formation radius of HCN derived for R\,Crt and IK\,Tau are sufficiently robust.

\end{appendix}

\end{document}